\documentclass[preprint,eqsecnum,aps,showpacs,floatfix]{revtex4}

\usepackage{graphicx}

\usepackage{amsmath,amssymb}

\usepackage{graphicx,epsfig}

\usepackage{dcolumn}
\usepackage{bm}

\begin{document}


\title{SUSY into Darkness: Heavy Scalars in the CMSSM}

\author{Van E. Mayes}

\affiliation{Department of Chemistry, The University of Texas at Tyler,
Tyler, TX 75799}

\begin{abstract}
A survey of the mSUGRA/CMSSM parameter space is presented.
The viable regions of the parameter space which satisfy 
standard experimental constraints are identified and discussed.
These constraints include a $124-127$~GeV mass 
for the lightest CP-even Higgs and the correct relic density for cold dark matter (CDM).    
The superpartner spectra corresponding to these regions fall within the well-known
hyperbolic branch (HB) and  
are found to possess sub-TeV neutralinos and charginos, with 
mixed Bino/Higgsino LSP's with $200-800$~GeV masses.   
In addition, the models possess $\sim 3-4$~TeV gluino masses and heavy 
squarks and sleptons with masses $m_{\tilde{q}}$, $m_{\tilde{l}}> m_{\tilde{g}}$.  
Spectra with a Higgs mass $m_h \cong 125$~GeV
and a relic density $0.105 \leq \Omega_{\chi^0} h^2 \leq 0.123$ are found to require EWFT 
at around the one-percent level, while those spectra with a much lower relic density require
EWFT of only a few percent.  
Moreover, the SI neutralino-proton direct detection
cross-sections are found to be 
below or within the XENON100 $2\sigma$ limit and should be experimentally accessible now 
or in the near future.
Finally, it is pointed-out that the supersymmetry breaking soft terms corresponding
to these regions of the mSUGRA/CMSSM parameter space 
($m_0\propto m_{1/2}$ with $m_0^2 >> m_{1/2}^2$ and $A_0 = -m_{1/2}$) 
may be obtained from general flux-induced 
soft terms in Type IIB flux compactifications with D3 branes.   
\end{abstract}


\maketitle
\thispagestyle{empty}

\newpage
\setcounter{page}{1}
\section{Introduction}
The recent discovery of a Higgs-like particle with a mass in the range
$124-127$~GeV is perhaps the single greatest development in high-energy
physics in recent memory~\cite{:2012gk,:2012gu}.  If this particle is indeed the Higgs scalar,
it not only represents the final piece of the Standard Model (SM), but can 
potentially open a window into the world beyond the SM.  However,
an important question
that must be answered is the problem of how such an elementary scalar 
remains so 
light against quantum corrections, an issue
known as the hierarchy problem.

An elegant solution to the hierarchy problem is supersymmetry (SUSY).  
One of the best motivated and most studied extensions of the Standard Model (SM)
is the incorporaton of SUSY into the Minimal Supersymmetric
Standard Model (MSSM).  However, nature itself is not so elegant since the 
superpartners have not been observed with 
the same masses as their SM counterparts, and so SUSY must
be a broken symmetry.  
Although the exact mechanism and scale at which SUSY is broken in nature should it exist
is not known, simple calculations suggest that the masses of the superpartners should have
$\mathcal{O}(1$~TeV$)$ masses if SUSY solves the hierarchy problem without requiring
any fine-tuning.  Moreover, it can be shown that there is an upper bound on the Higgs mass
in the MSSM, $m_h \lesssim 130$~GeV~\cite{Carena:2002es}, which is in nice accord with the Higgs-like resonance observed
at the LHC.  Moreover, in addition to providing a solution to the hierarchy problem, 
SUSY with R-parity imposed can provide a natural candidate 
for dark matter~\cite{Ellis:1982wr,Ellis:1983wd,Ellis:1983ew}. Finally, the apparent convergence of the gauge 
couplings when extrapolated to high energies is more precise when SUSY
is incorporated compared to the non-SUSY SM, consistent with the idea of
Grand Unification~\cite{Dimopoulos:1981yj, Ibanez:1981yh}.

Despite these many attractive features of SUSY, data from the the Large Hadron Collider (LHC)
has been infringing upon this rosy scenario as of late.  In particular, the LHC
has thus far failed to find any new particles beyond the SM.  Indeed, direct searches
for squarks and gluinos are pushing the mass limits for these particles into the 
TeV range~\cite{:2012rz,Aad:2012hm,:2012mfa,Aad:2011ib,Chatrchyan:2011zy}.   
Furthermore, to obtain a $\sim125$~GeV Higgs mass in the MSSM requires large
radiative corrections involving the top/stop sector, requiring large stop
squark masses $\mathcal{O}($TeV$)$ and/or large values of tan$\beta$.   
In spite of this, reports of the demise of SUSY are
greatly exaggerated.  Indeed, in some extended models it is possible 
to obtain a $125$~GeV Higgs while maintaining a light spectrum of 
superpartners~\cite{Li:2012qv,Ellwanger:2009dp}.

Perhaps the most-studied framework for supersymmetry breaking is 
minimal supergravity (mSUGRA), or equivalently the Constrained MSSM (CMSSM)~\cite{Chamseddine:1982jx, Ohta:1982wn, Hall:1983iz}.  
However, to obtain a sufficiently large Higgs mass in mSUGRA/CMSSM 
seemingly requires heavy squarks and sleptons which generically spoils the naturalness in which
the hierarchy problem is solved by introducing some amount of electroweak fine-tuning (EWFT).  One possible 
exception to this is the hyperbolic branch (HB)/focus point (FP) region of the mSUGRA/CMSSM parameter
space characterized by large $m_0$ in comparison to $m_{1/2}$ 
where the amount of required EWFT is minimized in respect to the full parameter 
space~\cite{Chan:1997bi,Feng:1999mn,Feng:1999zg,Baer:1995nq,Baer:1998sz,Chattopadhyay:2003xi}.  
Several different groups have recently reassessed the status of mSUGRA/CMSSM
in light of the $\sim 125$~GeV Higgs 
discovery~\cite{Kadastik:2011aa,Strege:2011pk,Aparicio:2012iw,Ellis:2012aa,Baer:2012uya,Matchev:2012vf,Akula:2012kk,Ghosh:2012dh,Fowlie:2012im,Buchmueller:2012hv,Strege:2012bt,Citron:2012fg,Ellis:2012nv,Boehm:2012rh} (see~\cite{Okada:2012nr} for a similiar analysis in the context of anomaly mediation). 
It is generally agreed that 
the mSUGRA/CMSSM parameter space is being squeezed by this discovery and pushed
into regions which require a degree of fine-tuning.  In~\cite{Baer:2012mv}, a study of the parameter space
in regards to fine-tuning was performed and it was concluded that there are no
regions where the Higgs is sufficiently heavy and where the relic density may
satisfy the WMAP constraint that do not require large fine-tuning.

In the following, scans of
the mSUGRA/CMSSM parameter space have been performed.  
Viable regions of the parameter space, which 
appear to fall within the HB region of the CMSSM parameter space, are identified.
In contrast to what was found~\cite{Baer:2012mv}, these regions do not seem to require excessive EWFT.  
The superpartner spectra corresponding to these regions
will be found to possess sub-TeV neutralinos and charginos, with 
mixed Bino/Higgsino LSP's with $200-800$~GeV masses. 
In addition, the models will be shown to possess $\sim 3-4$~TeV gluino masses and heavy 
squarks and sleptons with masses $m_{\tilde{q}}$, $m_{\tilde{l}}> m_{\tilde{g}}$.  
Spectra with a Higgs mass $m_h \geq 125$~GeV
and a relic density $0.105 \leq \Omega_{\chi^0} h^2 \leq 0.123$ are found to require EWFT 
at around the one-percent level, while those spectra with a relic density much lower require
EWFT of only a few percent.  
Moreover, the spin-independent neutralino-proton 
cross-sections for direct detection of dark matter for these spectra are below the 
XENON100 limit~\cite{Aprile:2011hi,Aprile:2012nq} 
and should be 
experimentally accessible in the near future.   
Finally, it is pointed-out that the supersymmetry breaking soft terms corresponding
to these regions of the mSUGRA/CMSSM parameter space 
($m_0\propto m_{1/2}$ with $m_0^2 >> m_{1/2}^2$ and $A_0 = -m_{1/2}$) 
may be obtained from general flux-induced 
soft terms in Type IIB flux compactifications with D3 branes.  

\section{Parameter Space}
  
The most studied model of supersymmetry breaking is minimal 
supergravity (mSUGRA), which arises from adopting the simplest ansatz for the K\a"ahler 
metric, treating all chiral superfields symmetrically.  In this framework,~~$\mathcal{N}=1$ 
supergravity is broken in a hidden sector which is communicated to the observable sector 
through gravitational interactions.  Such models are characterized by the following 
parameters: a universal scalar mass $m_0$, a universal gaugino mass $m_{1/2}$, the 
Higgsino mixing $\mu$-parameter, the Higgs bilinear $B$-parameter, a universal 
trilinear coupling $A_0$, and tan~$\beta$.  One then determines the $B$ and $|\mu|$ 
parameters by the minimization of the Higgs potential triggering 
REWSB~\cite{Ellis:1983bp,AlvarezGaume:1983gj}, with the sign 
of $\mu$ remaining undetermined.    The soft terms are then input into {\tt MicrOMEGAs 2.4.5}~\cite{Belanger:2010pz,Belanger:2008sj,Belanger:2006is} using
{\tt SuSpect 2.40}~\cite{SUSP} as a front end to evolve the soft terms down to the electroweak scale via 
the Renormalization Group Equations (RGEs) and then to 
calculate the corresponding relic neutralino density.  We take the top quark mass to be
$m_t = 173.2 \pm 0.9$~GeV~\cite{Lancaster:2011wr} and leave tan~$\beta$ as a free parameter, while $\mu$ is determined by
the requirement of REWSB. However, we do take $\mu > 0$ as suggested by the results of 
$g_{\mu}-2$ for the muon.  In analyzing the resulting data, we consider the following experimental constraints:

\begin{figure}
  \centering
	\includegraphics[width=1.0\textwidth]{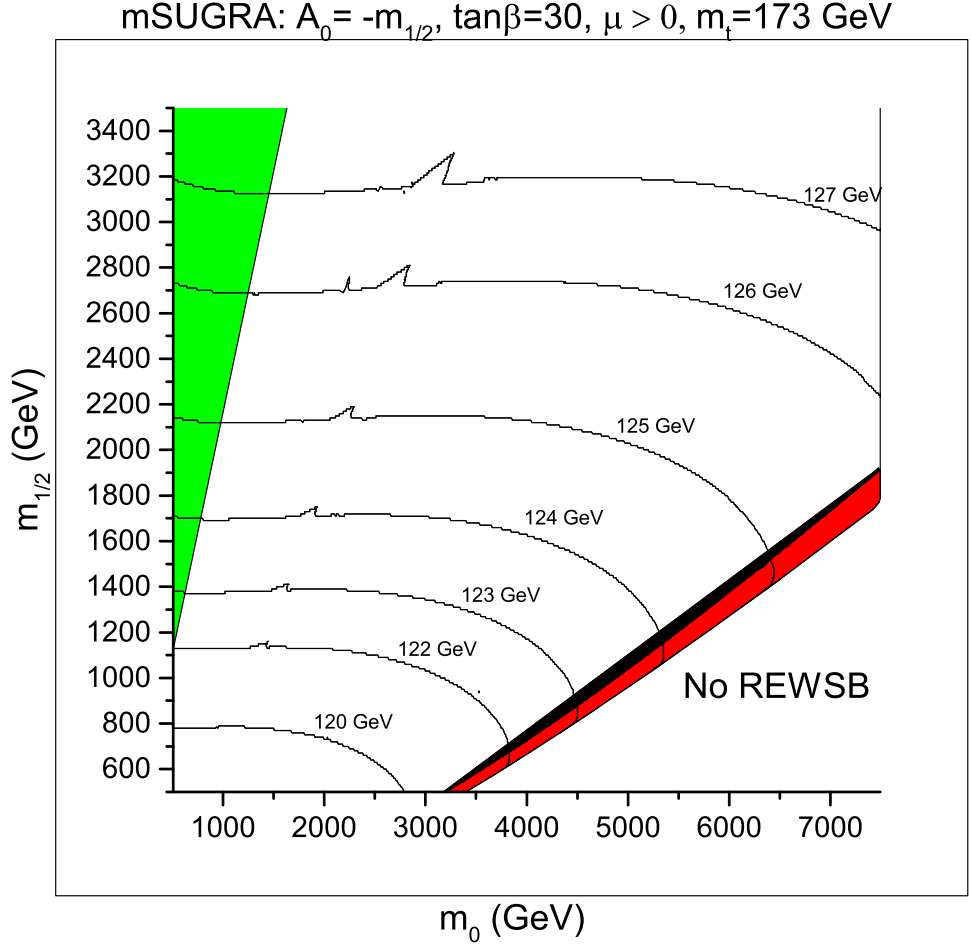}
		\caption{The mSUGRA $m_{1/2} \ vs.\ m_0$ plane with $A_0 = -m_{1/2}$, $\mu>0$, tan$\beta=30$, and $m_t=173$~GeV.  The region shaded in black indicates
		a relic density $0.105 \lesssim \Omega_{\chi^0} h^2 \lesssim 0.123$, the region shaded in red indicates $\Omega_{\chi^0} h^2 \lesssim 0.123$, while the region shaded in green has a charged LSP.  
		The black contour lines indicate the lightest CP-even Higgs mass.}
	\label{fig:mSUGRA_CountourPlanetb30}
\end{figure}

\begin{enumerate}

\item The WMAP 9-year $2-\sigma$ preferred range~\cite{Hinshaw:2012fq} for the cold dark matter density,  0.105 $\leq \Omega_{\chi^o} h^{2} \leq$ 0.123.   We consider two cases, one where a neutralino LSP is the dominant component of the dark matter and another where it makes up a subdominant component such that
0 $\leq \Omega_{\chi^o} h^{2} \leq$ 0.123\footnotetext[1]{The first results from the Planck experiment\cite{Ade:2013lta}, with a slightly larger value for the dark matter density $\Omega_c h^2 = 0.1199 \pm 0.0027$, appeared shortly after the first version of the paper was produced. Using the Planck result rather than the WMAP bounds results in a slight shifts in the parameter spaces shown in Figs.~[1-4], but does not alter the fundamental conclusions of this paper.}.

\item The experimental limits on the Flavor Changing Neutral Current (FCNC) process, $b \rightarrow s\gamma$. The results from the Heavy Flavor Averaging Group (HFAG)~\cite{HFAG}, in addition to the BABAR, Belle, and CLEO results, are: $Br(b \rightarrow s\gamma) = (355 \pm 24^{+9}_{-10} \pm 3) \times 10^{-6}$. There is also a more recent estimate~\cite{MMEA} of $Br(b \rightarrow s\gamma) = (3.15 \pm 0.23) \times 10^{-4}$. For our analysis, we use the limits $2.86 \times 10^{-4} \leq Br(b \rightarrow s\gamma) \leq 4.18 \times 10^{-4}$, where experimental and
theoretical errors are added in quadrature.

\item The process $B_{s}^{0} \rightarrow \mu^+ \mu^-$ which has recently been observed 
      to be in the range $2\times 10^{-9} < BF(B_{s}^{0} \rightarrow \mu^+ \mu^-) < 4.7\times 10^{-9}$ by LHCb~\cite{:2012ct}.

\item The lightest CP-even Higgs mass in the range $124$~GeV$\lesssim m_h \lesssim 127$~GeV as observed by the ATLAS and CMS experiments at the LHC~\cite{:2012gk,:2012gu}.

\end{enumerate}

In the following, we will not require that the anomalous magnetic moment of the muon~\cite{MUON}, $4.7\times 10^{-10} \leq a_{\mu} \approx 52.7\times 10^{-10}$, is solved by 
contributions from supersymmetric particles as the spectra that will be studied may only make a small contribution.  Furtheremore, there are large hadronic contributions
to this anomaly that require delicate subtractions with large uncertainties~\cite{Davier:2010nc}.

\begin{figure}
  \centering
  \begin{tabular}{cc}
	\includegraphics[width=0.5\textwidth]{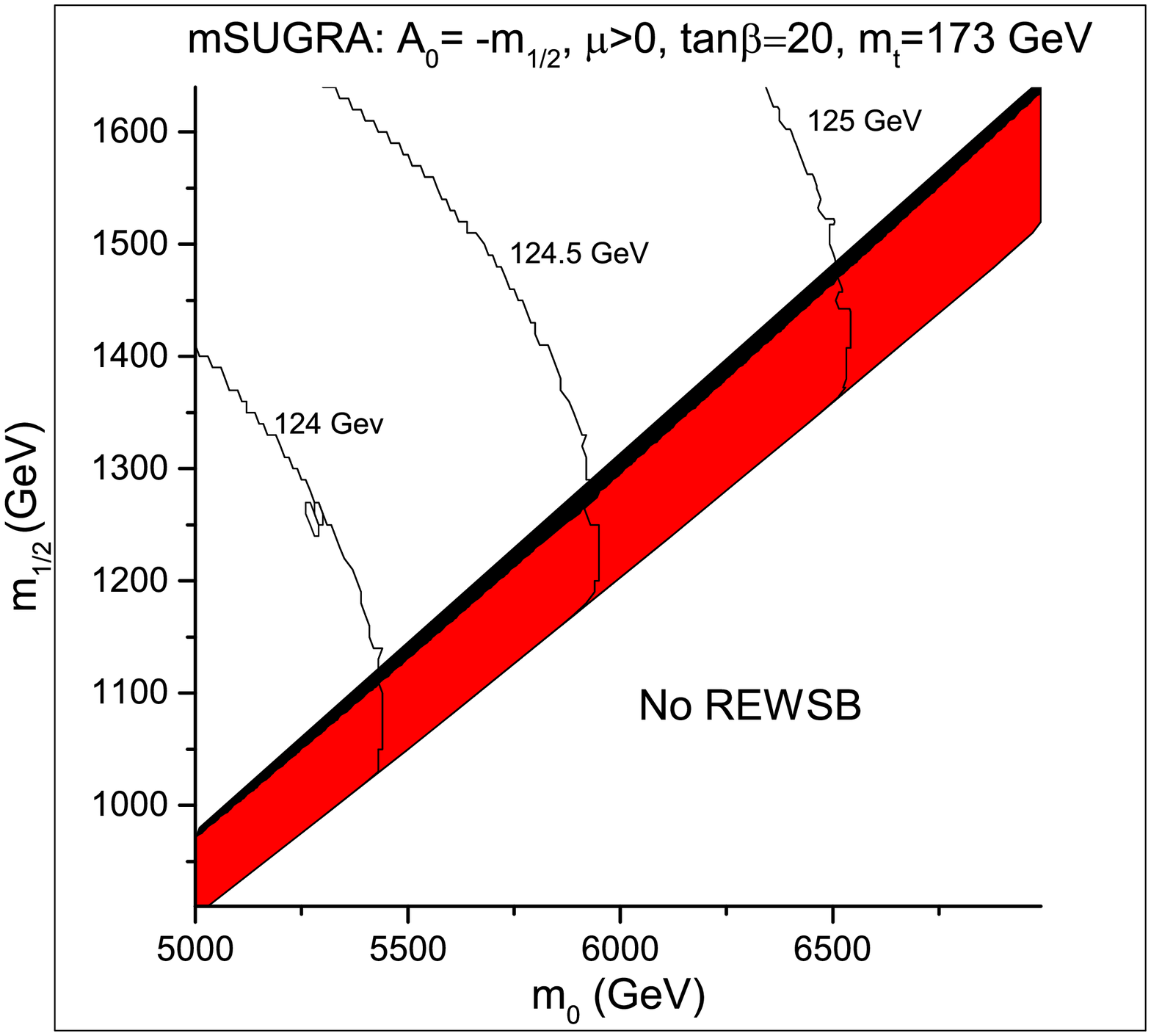}
	\includegraphics[width=0.5\textwidth]{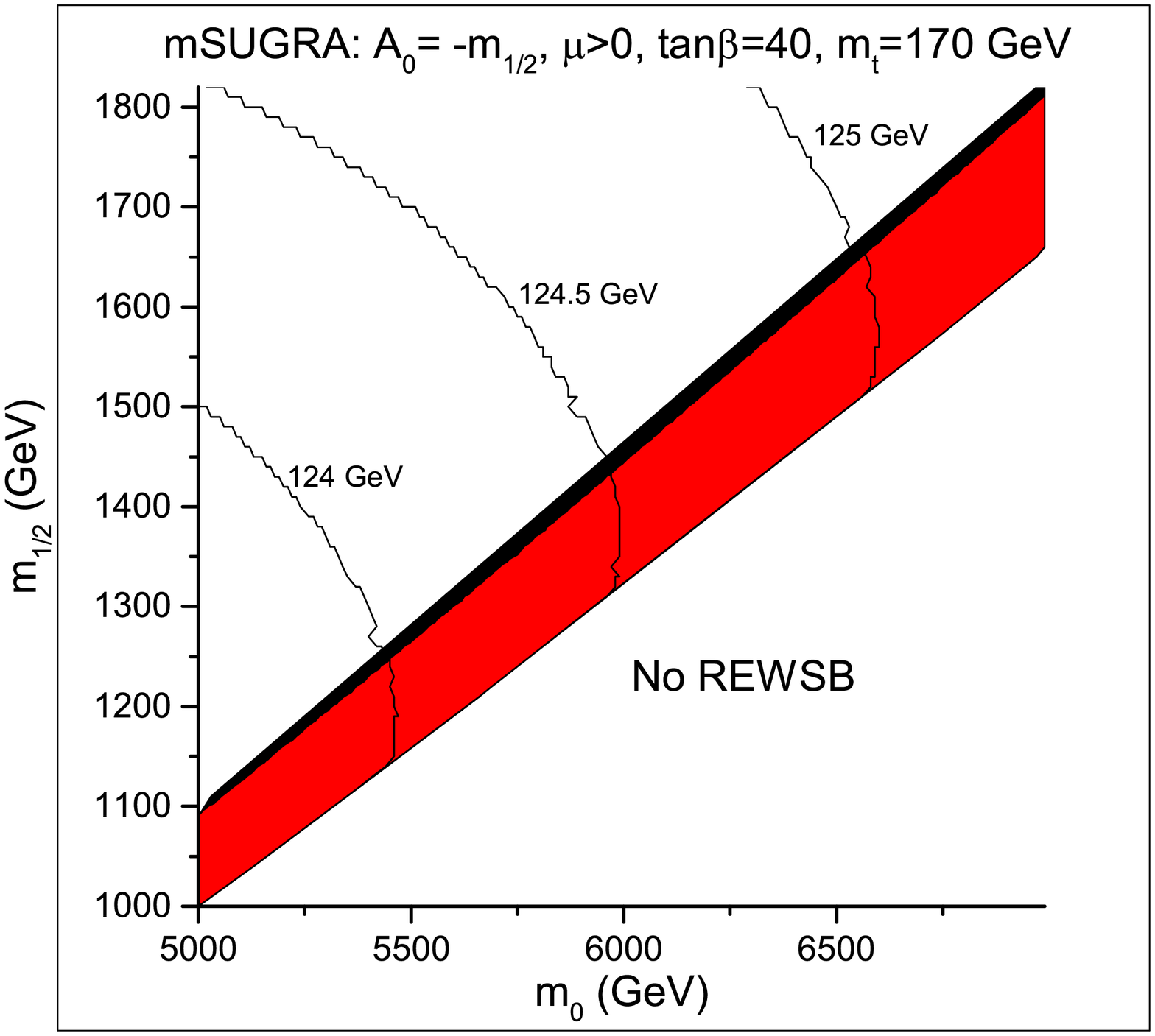}
	\end{tabular}
		\caption{The mSUGRA $m_{1/2} \ vs.\ m_0$ plane for tan$\beta=20$ and tan$\beta=40$ with $A_0 = -m_{1/2}$, $\mu>0$, and $m_t=173.1$~GeV.  The region shaded in black indicates
		a relic density $0.102 \lesssim \Omega_{\chi^0} h^2 \lesssim 0.123$, while the region shaded in red indicates $\Omega_{\chi^0} h^2 \lesssim 0.123$.  
		The black contour lines indicate the lightest CP-even Higgs mass.}
	\label{fig:mSUGRA_CountourPlanetb20}
\end{figure}

\begin{table}[t]
	\footnotesize
	\renewcommand{\arraystretch}{1.0}
	\begin{center}
	\caption{Low energy supersymmetric particles and their masses (in GeV) for $m_{1/2} = 1560$,
$m_{0} = 6510$, $A_{0} = -1560$, tan$\beta = 30$, $\mu > 0$ and $m_t=173.1$~GeV.  
Here $\Omega_{\chi^{o}} h^{2}$ =  0.103, $\sigma^{SI}_{p-\chi^0}=3.575\times10^{-8}$~pb, $\Delta_{EW}=111.4$,
$Br(B_s \rightarrow \mu^+\mu^-) = 3.05\times 10^{-9}$, $a_{\mu}=0.2575\times 10^{-10}$, and $Br(b \rightarrow s \gamma) = 3.2\times 10^{-4}$.}
	\begin{tabular}{|c|c|c|c|c|c|c|c|c|c|}\hline

$h^0$ & $H^0$ & $A^0$ & $H^{\pm}$ & ${\widetilde g}$ & $\widetilde\chi_1^{\pm}$
& $\widetilde\chi_2^{\pm}$ & $\widetilde\chi_1^{0}$ & $\widetilde\chi_2^{0}$  \\ \hline

125.1   & 5434.5   & 5434.5    & 5435.3   &  3636  & 691.1    & 1337   & 663.1
&  696.7      \\ \hline

$\widetilde\chi_3^{0}$ & $\widetilde\chi_4^{0}$ & ${\widetilde t}_1$ & ${\widetilde t}_2$ &
${\widetilde u}_R/{\widetilde c}_R$ & ${\widetilde u}_L/{\widetilde c}_L$ &
${\widetilde b}_1$ & ${\widetilde b}_2$ & \\ \hline

728.7 & 1337  & 4516  & 5657 & 7026 & 6997 & 5666 & 6481 &\\ \hline

${\widetilde d}_R/{\widetilde s}_R$ & ${\widetilde d}_L/{\widetilde s}_L$ &
${\widetilde \tau}_1$ & ${\widetilde \tau}_2$ & ${\widetilde \nu}_{\tau}$ &
${\widetilde e}_R/{\widetilde \mu}_R$ & ${\widetilde e}_L/{\widetilde \mu}_L$ &
${\widetilde \nu}_e/{\widetilde \nu}_{\mu}$ & $LSP$\\ \hline

7026 & 6995 & 6018 & 6312 & 6311 & 6556 &  6522 & 6556 & \textit{Bino/Higgsino}\\
	\hline
	\end{tabular}
	\label{tab:SUSYSpectrum1}
	\end{center}
\end{table}

\begin{table}[t]
	\footnotesize
	\renewcommand{\arraystretch}{1.0}
	\begin{center}
	\caption{Low energy supersymmetric particles and their masses (in GeV) for $m_{1/2} = 1910$,
$m_{0} = 7460$, $A_{0} = -1910$, tan$\beta = 30$, $\mu > 0$, and $m_t=173.1$~GeV.  
Here $\Omega_{\chi^{o}} h^{2}$ =  0.113, $\sigma^{SI}_{p-\chi^0}=3.089\times10^{-8}$~pb, $\Delta_{EW}=162.9$, 
$Br(B_s \rightarrow \mu^+\mu^-) = 3.06\times 10^{-9}$, $a_{\mu}=0.1920\times 10^{-10}$, and $Br(b \rightarrow s \gamma) = 3.2\times 10^{-4}$.}
	\begin{tabular}{|c|c|c|c|c|c|c|c|c|c|}\hline

$h^0$ & $H^0$ & $A^0$ & $H^{\pm}$ & ${\widetilde g}$ & $\widetilde\chi_1^{\pm}$
& $\widetilde\chi_2^{\pm}$ & $\widetilde\chi_1^{0}$ & $\widetilde\chi_2^{0}$  \\ \hline

125.8   & 6242.3  & 6242.3    & 6243.0   &  4368.0  & 837.3    & 1637.0   & 815.4
&  841.8      \\ \hline

$\widetilde\chi_3^{0}$ & $\widetilde\chi_4^{0}$ & ${\widetilde t}_1$ & ${\widetilde t}_2$ &
${\widetilde u}_R/{\widetilde c}_R$ & ${\widetilde u}_L/{\widetilde c}_L$ &
${\widetilde b}_1$ & ${\widetilde b}_2$ & \\ \hline

886.5 & 1637  & 5267  & 6569 & 8127 & 8086 & 6581 & 7496 &\\ \hline

${\widetilde d}_R/{\widetilde s}_R$ & ${\widetilde d}_L/{\widetilde s}_L$ &
${\widetilde \tau}_1$ & ${\widetilde \tau}_2$ & ${\widetilde \nu}_{\tau}$ &
${\widetilde e}_R/{\widetilde \mu}_R$ & ${\widetilde e}_L/{\widetilde \mu}_L$ &
${\widetilde \nu}_e/{\widetilde \nu}_{\mu}$ & $LSP$\\ \hline

8127 & 8082 & 6902 & 7246 & 7246 & 7525 &  7478 & 7524 & \textit{Bino/Higgsino}\\
	\hline
	\end{tabular}
	\label{tab:SUSYSpectrum2}
	\end{center}
\end{table}

\begin{figure}
  \centering
	\includegraphics[width=1.0\textwidth]{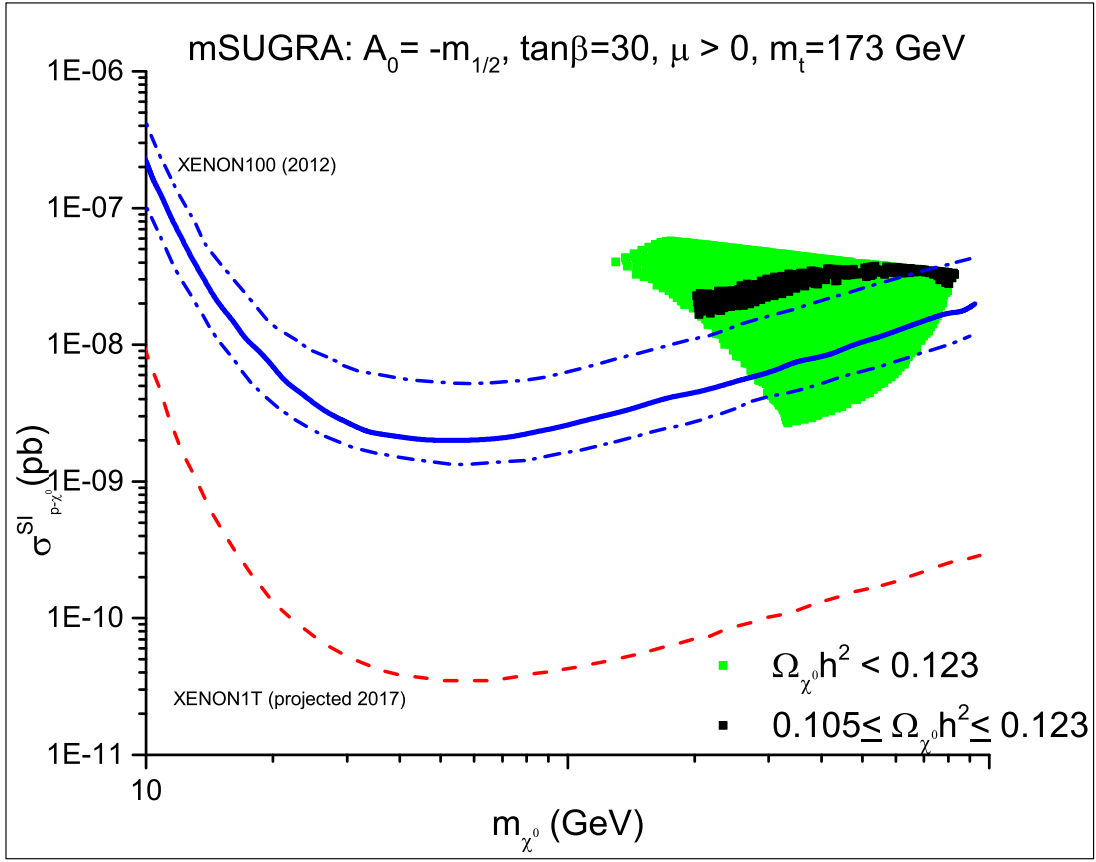}
		\caption{The spin-independent (SI) neutralino-proton direct detection cross-sections vs. neutralino mass for regions of the parameter space where $\Omega_{\chi^0}h^2\leq 0.123$. The region shaded
		in black indicates $0.105 \lesssim \Omega_{\chi^0}h^2 \lesssim 0.123$.  The upper limit on the cross-section obtained from the XENON100 experiment is shown in blue with the $\pm 2\sigma$ bounds shown as dashed curves, while the red dashed curved indicates the future reach of the XENON1T experiment.}
	\label{fig:DirectDetCrossSectionsvsNeutralinoMasstb30A}
\end{figure}

\begin{figure}
  \centering
	\includegraphics[width=1.0\textwidth]{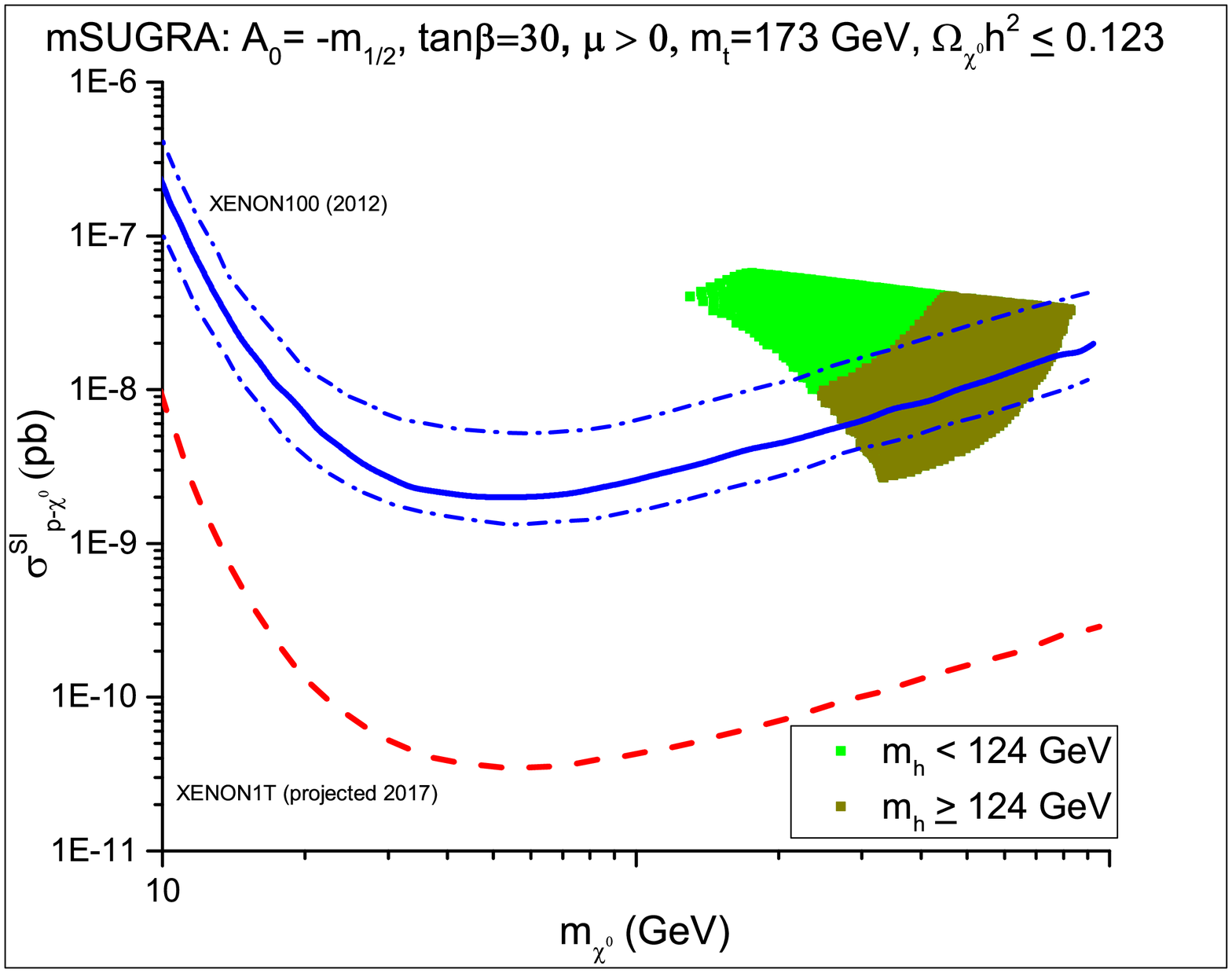}
		\caption{The spin-independent (SI) neutralino-proton direct detection cross-sections vs. neutralino mass for regions of the parameter space where $\Omega_{\chi^0}h^2\leq 0.123$. The region shaded
		in yellow-green a Higgs mass $m_h \geq 124$~GeV.  The upper limit on the cross-section obtained from the XENON100 experiment is shown in blue with the $\pm 2\sigma$ bounds shown as dashed curves, while the red dashed curve indicates the reach of the XENON1T experiment.}
	\label{fig:DirectDetCrossSectionsvsNeutralinoMasstb30B}
\end{figure}

In order to generate superpartner and Higgs spectra, we shall work within the 
mSUGRA/CMMSM framework.  Here we will generate a set of soft terms for the mSUGRA/CMSSM
parameter space.
We take the top quark mass to be $m_t = 173.2 \pm 0.9$~GeV.  
We vary $m_0$ and $m_{1/2}$ each in increments of $10$~GeV between 
$500-7500$~GeV for each scan.  In addition, we fix $A_0 = -m_{1/2}$ as this relation is typical of 
soft terms induced from fluxes in Type IIB string compactifications.  
Scans are made for different values of tan$\beta$, while $\mu$ is determined by the requirement of 
radiative electroweak symmetry breaking (EWSB).  In addition to imposing experimental constraints,
the spectra are filtered from the final data set if the iterative procedure employed  by {\tt SuSpect} 
does not converge to a reliable solution. 

A contour plot of the $m_{1/2} \ vs. \ m_{0}$ plane for tan$\beta=30$ is shown in Fig.~\ref{fig:mSUGRA_CountourPlanetb30}.  Regions satisfying different constraints are as indicated
on the figure.  Here, we can see that there is a linear band shaded in black where the lightest neutralino relic density satisfies 
$0.105 \lesssim \Omega_{\chi^0}h^2 \lesssim 0.123$ which sits inside a broader linear band shaded in red where the relic density statisfies $\Omega_{\chi^0}h^2 \lesssim 0.123$.  
These bands lie along the HB branch of the mSUGRA/CMSSM parameter space.  The values of the lightest CP-even Higgs mass are indicated
on the plot by the black contours lines.  It can be seen from this plot that there are regions of this parameter space where 
the relic density is in the range $0.105 \lesssim \Omega_{\chi^0}h^2 \lesssim 0.123$ and where the desired Higgs mass may be also obtained. 
Please note that although these plots seem to indicate that these spectra lie along a continuous band, they are actually
interspersed with spectra where {\tt SuSpect} is not able to converge to a solution.   
Sample spectra
with $m_h = 125.2$~GeV and $\Omega_{\chi^0} h^2 = 0.103$ are shown in 
Table~\ref{tab:SUSYSpectrum1}, and with $m_h = 125.8$~GeV and $\Omega_{\chi^0} h^2 = 0.113$  
is shown in Table~\ref{tab:SUSYSpectrum2}.  As is typical for spectra
in the HB region of the parameter space, the lightest neutralino has a large Higgsino component while the squarks and sleptons
all have masses greater than the gluino mass.  For all of the spectra for which the Higgs mass satisfies $124$~GeV$\lesssim m_h \lesssim 127$~GeV
and for which the relic density satisifes the WMAP constraint, the gluino mass is in the range $3-4$~TeV.  Thus, these spectra result in 
the \lq Higgsino World\rq \ scenario~\cite{Baer:2011ec}.  Due to the heavy masses for the gluino and squarks in these models, it would be
very difficult to observe any superpartners at the LHC if the spectrum of superpartners falls into these regions of the
parameter space.  However, the prospects for observing superpartners at a linear collider or at a higher-energy hadron collider 
appear to be more promising. Similar results are obtained for tan$\beta=20$ and tan$\beta=40$ as can be seen in Fig.~\ref{fig:mSUGRA_CountourPlanetb20}.

While these spectra may not create an observable signal at the LHC, the relic neutralino-proton SI cross-sections 
for dark matter direct detection are currently being probed by the XENON100 experiment.  Plots of the SI neutralino-proton
cross-sections vs. neutralino mass are shown in 
Fig.~\ref{fig:DirectDetCrossSectionsvsNeutralinoMasstb30A} and Fig.~\ref{fig:DirectDetCrossSectionsvsNeutralinoMasstb30B}.  
As can be seen from these plots, regions
of the parameter space with a Higgs mass $m_h \lesssim 124$~GeV and a relic density in the range 
$0.105 \lesssim \Omega_{\chi^0} h^2 \lesssim 0.123$ have been excluded by the upper limit on the proton-neutralino
SI cross-section from XENON100.  However, regions of the parameter space with $m_h \gtrsim 124$~GeV and 
a relic neutralino density at or below the WMAP limit are still viable, at least within the $2\sigma$ range.
These regions of the parameter 
space should either be excluded in the next update, or they should see a clear signal.  In particular,
the XENON1T experiment~\cite{Aprile:2012zx} should be able to completely probe this parameter space.  However, it should be noted
that the dark matter constraint on this parameter may only be imposed if R-parity is conserved. Thus, even 
if the XENON1T experiment reports negative results, the supersymmetry parameter space would still be viable
if R-parity violation is allowed.   

\section{Fine-Tuning}

\begin{figure}
  \centering
	\includegraphics[width=1.0\textwidth]{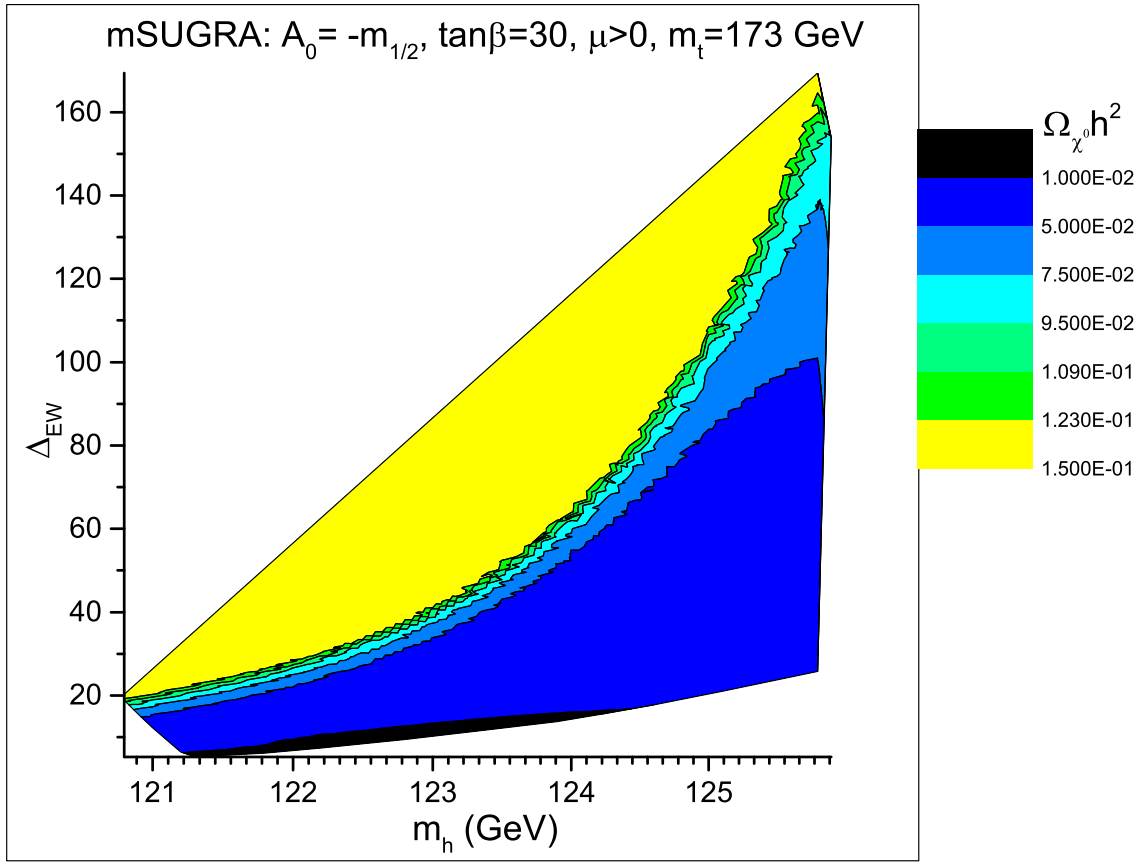}
		\caption{Contour plot of the $\Delta_{EW}$~vs.$~m_h$ for the parameter space with $\Omega_{\chi^0} h^2 \leq 0.15$. The different
		colors denote different ranges for the neutralino relic density, $\Omega_{\chi^0} h^2$. The areas covered in green indicate regions of 
		the parameter space with a relic density which falls into the WMAP preferred range.}
	\label{fig:mSUGRA_FineTuningvsHiggsMasstb30}
\end{figure}

One of the strongest reasons for introducing low-scale SUSY is to solve the hierarchy problem.  The parameter
space which has been found does this, however it is an important question whether or not this is
accomplished naturally without reintroducing any fine-tuning (the \textit{little hierarchy problem}).  Ordinarily, such spectra with large scalar 
masses would generically be considered fine-tuned.  This is not necessarily true for those spectra which fall 
in the HB region of the parameter space, such as those falling upon the red and black bands of Fig.~\ref{fig:mSUGRA_CountourPlanetb30}.
The amount of fine-tuning with respect to the electroweak scale (EWFT) is typically signified by the fine-tuning parameter 
\begin{equation}
\Delta_{EW} \equiv \mbox{max}(C_i)/(M_Z^2/2),
\end{equation}
where $C_{\mu}\equiv \left|-\mu^2\right|$, $C_{H_u}\equiv\left|-m^2_{H_u}tan^2\beta/(tan^2\beta-1)\right|$, and 
$C_{H_d}\equiv \left|-m^2_{H_d}/(tan^2\beta-1)\right|$.  The percent-level of EWFT is then given by $\Delta_{EW}^{-1}$.
It should be noted that for most of the parameter space explored in this analysis $C_{\mu}$ is dominant, 
and so generally we have
$\Delta_{EW}=\left|-\mu^2\right|/(M_Z^2/2)$.  In~\cite{Baer:2012mv}, it is argued that $\Delta_{EW}$ only provides a measure of
the minimum amount of fine-tuning in regards to the electroweak scale and provides no information about the high
scale physics involved in a particular model of SUSY breaking.  In order to provide a measure of how fine-tuned a particular
model is given knowledge of how SUSY is broken at a high energy scale, a parameter called $\Delta_{HS}$ was introduced which
is analogous to $\Delta_{EW}$~\cite{Baer:2012mv}.  For most of the parameter space this parameter is given by
\begin{equation}
\Delta_{HS} = \frac{m_0^2+\mu^2}{(M_Z^2/2)}= \Delta_{EW}+\frac{m_0^2}{(M_Z^2/2)}.
\end{equation}
As we can see, for regions of the mSUGRA/CMSSM parameter space with large scalar masses, $\Delta_{HS}$ is
very large even for those cases where $\Delta_{EW}$ is small such as in the HB regions.  This simply reflects
the fact that, although a particular SUSY spectrum may be completely natural and solve the hierarchy problem
without any fine-tuning, obtaining this spectrum within the mSUGRA/CMSSM framework of SUSY breaking 
requires large cancellations which only happens for specific sets of soft-terms rather than the general parameter 
space.  

In Fig.~\ref{fig:mSUGRA_FineTuningvsHiggsMasstb30}, a contour plot of $\Delta_{EW}$~vs.~$m_h$ is shown for the parameter
space satisfying $\Omega_{\chi^0} h^2 \leq 0.123$.  The different colored regions of the plot
denote different ranges of $\Omega_{\chi^0} h^2$.  From this plot, it can be seen that the amount of EWFT appears 
to be proportional to $m^2_{H_u} - m^2_{H_d}$ as might be expected from the Higgs potential.  
In addition, the amount of EWFT seems to increase
linearly with relic density.  For spectra with a relic density $0.105 \leq \Omega_{\chi^0} h^2 \leq 0.123$ and 
a Higgs mass $m_h=124$~GeV, $\Delta_{EW}\approx 60$ and thus requires minimum EWFT at about the two-percent level.  
On the other hand, spectra with the same relic density and a Higgs mass $m_h=125$~GeV requires minimum EWFT 
at the percent level.  Conversley, spectra with a very low relic density can have a $125$~GeV Higgs mass
and only require EWFT at the five-percent level.  However, in this case the neutralino LSP can only provide
a small component of the cold dark matter (see~\cite{Amsel:2011at} for a similar study prior to the discovery 
Higgs-like resonance).

\section{Flux-induced Soft Terms on D3 branes}

From the analysis of the previous sections, it can be seen that there are spectra
within the mSUGRA/CMSSM parameter space which may solve the hierarchy problem while only
requiring EWFT of a percent or greater.  However, as discussed in the last section
it does require a large amount of fine-tuning to obtain these spectra within 
the specific framework of supersymmetry breaking, mSUGRA/CMSSM.  Within this framework,
it is rather unnatural to have a universal scalar mass which is so much larger 
than the universal gaugino mass, $m_0^2 >> m_{1/2}^2$, as large cancellations are
required to obtain a light Higgs mass.  Clearly, it would be desirable to have a
specific model of supersymmetry breaking for which large scalar masses in comparison
to the gaugino masses arise naturally.  

Over the past decade, there has been much progress in constructing realistic 
models in Type I and Type II string compactifications~\cite{Blumenhagen:2005mu, Blumenhagen:2006ci}.  
In these models,
the SM fields are localized within the world-volume of D-branes embedded
in a closed 10-d closed string background.  Physical observables such as 
gauge and Yukawa couplings are dependent upon the moduli of compactification,
which must be stabilized in order to have a true vaccum. 
It has been shown that in Type IIB compactifications 
non-trivial backgrounds
of NSNS and RR 3-form field strength fluxes generically fix the VEVs of
the dilaton and all complex structure moduli.  

Besides fixing the VEVs of the
moduli fields, these fluxes may also induce SUSY-breaking soft terms.  
As shown in~\cite{Camara:2003ku}, for the most general combination involving both
imaginary selfdual (ISD) and imaginary anti-selfdual (IASD) fluxes,
the soft terms on D3 branes take the form
\begin{eqnarray}
&m_0^2 = \frac{\left|m_{1/2}\right|^2}{3}\left[1-\mbox{tan}\theta~\mbox{cos}(\delta + \beta)\right], \\ \nonumber
&A^{ijk} = -m_{1/2}h^{ijk}.
\end{eqnarray}
For real flux backgrounds ($\delta=\beta=0$~mod~$2\pi$), and tan$\theta=0$ the flux-induced soft terms 
take the dilaton-dominated form of no-scale supergravity.
In addition, for tan$\theta >> -1$ one
finds that $m_0 \propto m_{1/2}$ with $m_0 >> m_{1/2}$ and $A= -m_{1/2}$, which is exactly the form
of the soft terms required to match the viable region of the parameter space found in the analysis of the previous 
sections. 

Thus, if the MSSM is built on D3 branes in Type IIB string theory with a combination of ISD and IASD
fluxes in the background, then it is possible to induce soft terms of the form studied in
this paper.  In such a model of SUSY-breaking, large scalar masses with $m_0 >> m_{1/2}$ are then 
completely natural in constrast to the situation with mSUGRA/CMSSM where there is no {\it a priori} correlation
between $m_0$ and $m_{1/2}$  
\footnotetext[2]{However, there is one problem with this scenario.  It is known that only ISD fluxes solve the equations 
of motion.  Thus, it is only possible to stabilize the moduli with ISD fluxes.  With only D3 branes and
including both ISD and IASD fluxes, one may have soft terms of the desired form, but one would require
other nonperturbative effects to stabilize the moduli.}.

\section{Conclusion}

We have surveyed the mSUGRA/CMSSM parameter space for tan$\beta=20$, tan$\beta=30$, and tan$\beta=40$
with the restriction $A_0= -m_{1/2}$.  
We have found that there are viable areas of the parameter space where the lightest CP-even Higgs mass is
in the range $124$~GeV $\lesssim m_h \lesssim 127$~GeV, the relic neutralino density is below the WMAP
constraint, $\Omega_{\chi^0}h^2 \leq 0.123$, and standard experimental constraints are satisfied.  These areas
of the parameter space appear to lie along the HB regions of the mSUGRA/CMSSM parameter space.  The corresponding 
spectra features neutralinos and charginos with sub-TeV masses, gluino masses in the range $3-4$~TeV, and 
heavy squarks and sleptons with masses greater than $4$~TeV.  The lightest neutralino is a mixture of 
Bino and Higgsino.  It would be difficult for these spectra to produce an observable signal at the LHC.  However, the prospects
for their observation at a linear collider are much more promising.  

While these spectra may not create an observable signal at the LHC, the relic neutralino-proton SI cross-sections 
for dark matter direct detection are currently being probed by the XENON100 experiment.  At present, regions
of the parameter space with a Higgs mass $m_h \lesssim 124$~GeV and a relic density in the range 
$0.095 \lesssim \Omega_{\chi^0} h^2 \lesssim 0.125$ have been excluded by the upper limit on the proton-neutralino
SI cross-section from XENON100.  However, regions of the parameter space with $m_h \gtrsim 124$~GeV and 
a relic neutralino density at or below the WMAP limit are still viable.  These regions of the parameter 
space should either be excluded in the next update, or they should see a discernable signal.  

We have also investigated the question of fine-tuning with respect to both the electroweak scale
and the high scale of supersymmetry breaking.  We have found that the spectra with a large enough Higgs mass 
$124$~GeV $\lesssim m_h \lesssim 127$~GeV
and the correct relic density $0.109 \lesssim \Omega_{\chi^0} h^2 \lesssim 0.123$
require at least a one-percent EWFT, while spectra satifying the Higgs constraint but 
which possess a low relic density are fine-tuned at the five-percent level.  As these spectra
fall into regions of the parameter space where $m_0^2 \ggg m_Z^2$, these spectra are highly fine-tuned
with respect to the high scale, at least within the context of mSUGRA/CMSSM.  

Finally, we have discussed the inducement of SUSY-breaking soft-terms from supergravity fluxes 
which appear in Type IIB string compactifications.  Such fluxes may be utilized in regards
to the moduli stabilization problem of string theory compactifications.  
We have pointed out that for a general combination of ISD and IASD
fluxes with D3 branes, the soft terms may have exactly the same form as those which give rise to the 
viable parameter
space we have investigated in the paper, namely $m_{0}\propto m_{1/2}$ with $m_0 \gg m_{1/2}$ and 
$A_0 = -m_{1/2}$.  Thus, in contrast to mSUGRA/CMSSM, having large $m_0$ compared to $m_{1/2}$ can
potentially arise naturally within the context of Type IIB flux compactifications, in constrast to the
situation with mSUGRA/CMSSM.

\section{Acknowledgments}
The author would like to thank The University of Texas at Tyler for providing resources which
allowed the production of this paper to be possible, as well as D.~V.~Nanopoulos and J.~Maxin 
for a critical reading of the manuscript.

\end{document}